\documentclass[12pt]{book}

\usepackage[dvips]{graphicx,color}
\usepackage{makeidx,tsukuba}

\makeauthorindex
\makeindex

\newcommand\paperno{
   \vspace{-8\baselineskip}
   \noindent \underline{\it LANL Report \rm \# LA-UR-03-3405}
   \vspace{6.8\baselineskip}}

\begin{document}

\BookTitle{\itshape The 28th International Cosmic Ray Conference}
\CopyRight{\copyright 2003 by Universal Academy Press, Inc.}
\pagenumbering{arabic}

\chapter{
Antiprotons in CR: What Do They Tell Us?}
\paperno

\author{%
%
%
I.V.~Moskalenko,$^{1,2}$ A.W.~Strong,$^3$ S.G.~Mashnik,$^4$ J.F.~Ormes$^1$\\
{\it 
(1) NASA/Goddard Space Flight Center, Code 661, Greenbelt, MD 20771, USA\\
(2) JCA/University of Maryland, Baltimore County, Baltimore, MD 21250, USA\\
(3) MPI f\"ur extraterrestrische Physik, Postfach 1312, 85741 Garching, Germany\\
(4) Los Alamos National Laboratory, Los Alamos, NM 87545, USA
} \\
}

\frenchspacing
\section*{Abstract}
Recent measurements of the CR $\bar p$ flux have been shown to pose
a problem for conventional propagation models.
In particular, models consistent with secondary/primary
nuclei ratio in CR produce too few $\bar p$'s, while matching the
ratio and the $\bar p$ flux requires \emph{ad hoc} assumptions.
This may indicate an additional local CR component or new phenomena in
CR propagation in the Galaxy.  We discuss several possibilities
which may cause this problem.

\section{Introduction}
The spectrum and origin of $\bar p$'s in CR is of interest for
many studies in physics and astrophysics.
Most of the observed CR $\bar p$'s are ``secondaries'' 
produced in collisions of CR particles with interstellar gas.
Their spectrum with a peak at about 2 GeV is distinctly different 
from other CR species.
Some proportion of the $\bar p$'s might originate in WIMP
annihilations and/or primordial black hole evaporation contributing 
mostly at low energies.
During the last decade there have been a number of space and balloon
CR experiments with improved sensivity and statistics, which 
impose stricter constraints on the 
models of CR propagation and heliospheric modulation.
It has been recently shown [5] that accurate $\bar p$
measurements during the last solar minimum 1995-97
[8] indicate a discrepancy
with calculations made using existing propagation models.
Because of the specific shape of the secondary $\bar p$
spectrum and the fact that their production in the ISM
can be calculated accurately,
$\bar p$'s provide a unique opportunity to test
models of CR propagation and heliospheric modulation.

\section{The Problem}
Secondary $\bar p$'s, $e^+$'s, and some proportion of diffuse 
$\gamma$-rays are products of interactions of mostly CR protons and 
He nuclei with interstellar gas. In
very much the same way spallation of CR nuclei 
give rise to secondary isotopes. Propagation of all particles
is governed by the same mechanism. The main constituents
are diffusion, energy losses and gains, particle production and 
disintegration. Because the mechanism is the same for \emph{all} particles,
a correct CR propagation model that describes any
secondary to primary ratio should equally well describe all the others: B/C,
sub-Fe/Fe, $\bar p/p$ ratios, and spectra of nuclei, $e^+$'s, and
diffuse continuum $\gamma$-rays.

The diffusive reacceleration models have certain 
advantages compared to other CR propagation models: they naturally reproduce
secondary/primary nuclei ratios in CR, have only three
free parameters (normalization and index of the diffusion
coefficient, and the Alfv\`en speed), and agree better with
K-capture parent/daughter nuclei ratio [3].
However, the reacceleration models designed to match the nuclei
ratios produce too few $\bar p$'s [5] because
matching the B/C ratio at all energies requires the diffusion
coefficient to be too large. 
The discrepancy is $\sim$40\% at 2 GeV while the stated
uncertainty in measured $\bar p$ flux in this energy range is now 
$\sim$20\%.
The conventional models without reacceleration based on 
local CR measurements, with simple energy 
dependence of the diffusion coefficient, and with uniform CR
source spectra throughout the Galaxy also fail to reproduce
simultaneously both the B/C ratio and $\bar p$ flux.

The difficulty associated with $\bar p$'s may indicate new effects,
if new experiments can confirm the BESS measurements.

\section{Discussion of Uncertainties}
The sources of uncertainties
which may affect the interpretation of $\bar p$ measurements
appear to be fourfold: (i)
propagation models and parameters, (ii) heliospheric modulation,
(iii) production cross sections of secondary nuclei
and $\bar p$'s, and (iv) systematic measurement errors.

(i) To this category we attribute errors in the 
Galactic gas distribution, ambient spectrum of CR, and our current
knowledge of CR diffusion process.
$\diamondsuit$
The errors in the gas distribution appear not to be 
so important in the
case of stable and long-lived nuclei. Such errors are compensated
simultaneously for all species
by the corresponding adjustment of the propagation parameters.
$\diamondsuit$
The local interstellar CR spectrum is studied 
quite well by direct
measurements at HE where solar modulation effects are
minimal, while the ambient CR proton spectrum on the large scale
remains unknown. 
The most direct test is provided by diffuse $\gamma$-rays,
but here we have a well known puzzle of the GeV excess 
in the EGRET data [2]. A possible explanation is the
inverse Compton scattering (ICS) of electrons whose
Galactic spectrum may be harder than the local one [9]; an 
explanation justified by the large electron energy losses. 
$\diamondsuit$
Our understanding of the CR propagation in the Galaxy is quite
basic. The distribution of CR sources is uneven
in space and random in time. The diffusion is assumed \emph{ad hoc} to be
governed by one unique mechanism over the decades of energy
MeV-PeV, while the diffusion coefficient is often taken
the same for the whole Galaxy. It is certainly an approximation,
but in fact most of CR data (except $\bar p$'s) support it.

(ii) Heliospheric modulation may introduce some error; it will
be similar for all CR nuclei which have
the charge/mass ratio about $+1/2$, except (anti-) protons
which have $\pm1$. Besides,
solar modulation for $\bar p$'s is different from that of protons,
due to charge sign dependent drift effects in the heliosphere.
At present,
several spacecraft provide information about particle fluxes
at different heliolatitudes (Ulysses) and close to the heliospheric 
boundary (Voyagers), which make the likelihood of a serious error small.
However, if modulation is weaker than assumed
a reacceleration model combining B/C, $\bar p$'s,
and other CR species is feasible.

(iii) Nuclear cross section errors are one of the main concerns. 
$\diamondsuit$
Fitting
B/C ratio in CR is a standard procedure to derive
the propagation parameters, while the calculated ratio, in turn, depends on
the total interaction and isotope
production cross sections. The latter have large
uncertainties, typically 
$\hbox{\rlap{\hbox{\lower3pt\hbox{$\sim$}}}\lower-2pt\hbox{$>$}}$20\%,
and sometimes larger.
In our calculations we use our own \emph{fits to the data} on cross sections 
$p+\mathrm{C,N,O} \to \mathrm{Be,B}$, that produce most of the Be and
B (see [7]). We thus can rule out a possibility of large errors in the
calculated B/C ratio.
$\diamondsuit$
Antiproton production cross section in $pp$-interactions is studied
quite well, while $\bar p$ production on nuclei relies on 
scarce data. This may lead to underestimation of the  
atmospheric contribution to the $\bar p$ flux measured in the upper atmosphere.
The flux of CR $\bar p$'s thus may be 
\emph{lower} at the top of the atmosphere by $\sim$25--30\% [1],
giving better agreement with our calculations in the reacceleration model.
An analysis of different parametrizations of $\bar p$
production on nuclei is given in [4].

(iv) Systematic measurement errors are difficult to account for,
but their effect can be reduced by careful choice of the data. 
The $\bar p$ data we rely on is the
flux at maximum $\sim$2 GeV where the statistical errors are minimal.
The spectra of protons and He are measured almost simultaneously and quite
precisely by BESS and AMS (see [5]). They also
agree with earlier experiments within
the error bars. The most accurate measurements of nuclei at low
energies are made by ACE, Ulysses, and Voyager and the agreement is
good. At HE the data obtained by HEAO-3 are the most
accurate and generally agree with earlier measurements;
we compare with the data in the middle of the interval where the systematic
errors should be minimal. However, new HE CR measurements are desirable.

\section{Alternative Possibilities}
A solution in terms of propagation models requires a break in the 
diffusion coefficient at a few GV [5]. It has been interpreted as 
change in the propagation mode; propagation of LE particles
may be aligned to the magnetic field lines rather than scattering. The chaotic
distribution of the magnetic field gives it a diffusion-like character.

If our local environment influences the spectrum of CR, then 
it is possible to solve the problem by
invoking a fresh ``unprocessed'' nuclei component
at LE [6], which may be produced in the Local Bubble. 
The idea is that primary CR like C and O have a local LE component, 
while secondary CR like B are produced Galaxy-wide over
the confinement time of 10--100 Myr. 
In this way an excess of B, which appears when propagation
parameters are tuned to match the $\bar p$ data, can be
eliminated by an additional local C (and the reduced 
Galactic production of B). 
The model appears to be able to describe a variety of
CR data, but at the price of additional parameters.

A consistent $\bar p$ flux in reacceleration
models can be obtained if there are sources of LE protons
$\hbox{\rlap{\hbox{\lower3pt\hbox{$\sim$}}}\lower-2pt\hbox{$<$}}$20 GeV. 
This energy is above the
$\bar p$ production threshold and effectively produces $\bar p$'s
at $\hbox{\rlap{\hbox{\lower3pt\hbox{$\sim$}}}\lower-2pt\hbox{$<$}}$2 GeV.  
The intensity and spectral shape of this
component could be derived by combining restrictions from $\bar p$'s
and diffuse $\gamma$-rays. 
This kind of nucleon spectrum was used in our HEMN model [9] to match the
spectrum of diffuse $\gamma$-rays as observed by EGRET [2]. 

More $\bar p$'s may be produced if
there is a population of
hard-nucleon-spectrum sources in the inner Galaxy.
Such a population is required since the ICS of hard-spectrum
electrons is insufficient to obtain an acceptable fit to the 
diffuse $\gamma$-ray latitude profiles [10].
Antiprotons produced by freshly accelerated particles in matter 
near the source can add up to $\bar p$'s produced Galaxy-wide.
We are going to address this possibility in future.


\emph{To summarize,} it is clear that accurate measurements of $\bar p$ flux
are the key to testing current
propagation models.
If new measurements confirm the $\bar p$ ``excess,'' current
propagation and/or modulation models will face a challenge.
If not -- it will be evidence that reacceleration model is
currently the best one to describe the data.

I.V.M.\ and S.G.M.\ acknowledge partial support from
a NASA Astrophysics Theory Program grant
and from the US Department of Energy (S.G.M.).

\section{References}
\re
1.\ Huang C.Y., Derome L., Bu\'enerd M.\ 2001, in Proc.\ 27th ICRC (Hamburg), 1707
\re
2.\ Hunter S.D.\ et al.\ 1997, ApJ 481, 205
\re 
3.\ Jones F.C., Lukasiak A., Ptuskin V., Webber W.\ 2001, in Proc.\ 27th ICRC (Hamburg), 1844
\re
4.\ Mashnik S.G., Moskalenko I.V.\ 2003, LANL preprint LA-UR-03-1610
\re
5.\ Moskalenko I.V., Strong A.W., Ormes J.F., Potgieter M.S.\ 2002, ApJ 565, 280
\re
6.\ Moskalenko I.V., Strong A.W., Mashnik S.G., Ormes J.F.\ 2003, ApJ 586, 1050
\re
7.\ Moskalenko I.V., Mashnik S.G.\ 2003, these Proc.
\re
8.\ Orito S.\ et al.\ 2000, Phys.\ Rev.\ Lett.\ 84, 1078
\re
9.\ Strong A.W., Moskalenko I.V., Reimer O.\ 2000, ApJ 537, 763 (err.\ 541, 1109)
\re
10.\ Strong A.W., Moskalenko I.V., Reimer O.\ 2003, these Proc.

\endofpaper
\end{document}